\documentclass[nofootinbib,twocolumn,preprintnumbers,superscriptaddress]{revtex4-1}
\usepackage{amsmath}
\usepackage{amssymb}
\usepackage{graphicx}
\usepackage{hyperref}
\usepackage{xspace}
\usepackage{bm}
\usepackage{slashed}
\usepackage{feynmp}

\usepackage{booktabs}
\AtBeginDocument{
\heavyrulewidth=.08em
\lightrulewidth=.05em
\cmidrulewidth=.03em
\belowrulesep=.65ex
\belowbottomsep=0pt
\aboverulesep=.4ex
\abovetopsep=0pt
\cmidrulesep=\doublerulesep
\cmidrulekern=.5em
\defaultaddspace=.5em
}

\newcommand{\be}{\begin{equation}}
\newcommand{\ee}{\end{equation}}
\newcommand{\ba}{\begin{eqnarray}}
\newcommand{\ea}{\end{eqnarray}}

\usepackage{xcolor}
\definecolor{light-gray}{gray}{0.8}

\begin{document}

\title{A note on the fate of the Landau-Yang theorem in non-Abelian gauge theories}

\preprint{CERN-PH-TH/2015-173}

\newcommand{\CERNaff}{CERN, PH-TH, CH-1211 Geneva 23, Switzerland}
\newcommand{\DIDaff}{Universit\'e Paris Diderot, F-75013 Paris, France}
\newcommand{\CNRSaff}{CNRS, UMR 7589, LPTHE, F-75005 Paris, France}
\newcommand{\SORBaff}{Sorbonne Universit\'es, UPMC Univ Paris 06, UMR
  7589, LPTHE, F-75005 Paris, France}
\newcommand{\sapienza}{Dipartimento di Fisica and  INFN, `Sapienza' Universit\`a di Roma, Piazzale Aldo Moro 2, I-00185 Roma, Italy.}
\newcommand{\edimburgo}{Higgs Centre for Theoretical Physics, School of Physics \& Astronomy,
  University of Edinburgh, Edinburgh EH9 3FD, United Kingdom.}
\newcommand{\ICREA}{ICREA, Instituci\'o Catalana de Recerca i Estudis Avan\c{c}ats, Barcelona, Spain}
\newcommand{\IFAE}{IFAE, Universitat Aut{\`o}noma de Barcelona, 08193 Bellaterra, Barcelona, Spain}

\author{Matteo Cacciari}
\affiliation{\DIDaff}
\affiliation{\SORBaff}
\affiliation{\CNRSaff}
\affiliation{\CERNaff}
\author{Luigi Del Debbio}
\affiliation{\edimburgo}
\affiliation{\CERNaff}
\author{Jos\'e R. Espinosa}
\affiliation{\ICREA}
\affiliation{\IFAE}
\author{Antonio D. Polosa}
\affiliation{\sapienza}
\author{Massimo Testa}
\affiliation{\sapienza}

\begin{abstract} 
   Using elementary considerations of Lorentz invariance, Bose symmetry and BRST  invariance,  we argue why the decay of a massive color-octet vector state into a pair of on-shell massless gluons is possible in a non-Abelian SU($N$) Yang-Mills theory, we constrain the form of the amplitude of the process and offer a simple understanding of these results in terms of effective-action operators.
\end{abstract}


\maketitle

\color{black}
\section{Introduction}

The Landau-Yang theorem states that a massive vector (i.e. spin 1) 
particle cannot decay into
two on-shell massless photons. The proofs of Landau~\cite{Landau:1948kw} and
Yang~\cite{Yang:1950rg} show that one can reach this conclusion under very
general conditions, using only Lorentz invariance, gauge invariance (in the form
of transversality of photon polarization vectors) and the Bose symmetry of the
photons.

One can also consider the case of a massive color-octet vector
state decaying into two on-shell massless gluons. This process is of phenomenological interest not only for models predicting the existence of colored massive vector particles, but also for heavy quarkonium physics, e.g. for the hadroproduction of a $J\!/\!\psi$ particle or its decay into hadrons.\footnote{In both cases the two gluons will not be exactly on-shell, but their off-shellness will be at most of the order of a few hundred MeV, and its effects therefore suppressed by the ratio with the much larger scale set by the $J\!/\!\psi$ mass or transverse mass.}
Evidence that this amplitude vanishes at tree-level has been given many times in
Quantum Chromodynamics (QCD), usually by explicitly calculating  the two-gluon decay of
a quark-antiquark pair projected onto a massive color-octet spin-1 state $(Q\bar
Q)_{\bm 1, \bm 8}$, see
e.g. 
Refs.~\cite{Cho:1995vh,Tang:1995zp,
Ma:1995fd,Tang:1996rm,Petrelli:1997ge,Beenakker:2013mva,Baernreuther:2013caa}.
In these papers it was generally understood, albeit with some exceptions, that the 
proofs of Landau and Yang could not be immediately extended to the color-octet case,
because of additional terms in the amplitude induced by 
the antisymmetric character of the color quantum numbers. However, for a number of years no attempt was apparently made to study the color-octet case in more detail or to explicitly check if the vanishing of the amplitude at tree level still held at higher orders.

This situation began to change only recently, when two calculations found non-zero results for the one-loop amplitude of the transition between a massive color-octet vector state and two massless gluons: Ref.~\cite{Chivukula:2013xla} calculated the next-to-leading order corrections to gluon fusion production of massive color-octet vector bosons (colorons), while Ref. \cite{Ma:2014oha} reported that the $(Q\bar Q)_{\bm 8}\to gg$ amplitude is also different from zero at one-loop level.

These results indicate that a colored version of the Landau-Yang theorem breaks down
once quantum corrections are taken into account. 
In this note we analyze in depth the origin of the cancellation at tree level and the structure of the amplitude in full generality, we explore where the proof of a would-be Landau-Yang theorem for colored states fails, and derive constraints on the form  of the amplitude for the decay of a massive vector color-octet in two massless gluons.
We also identify higher-dimension effective operators whose presence can explain the non-vanishing results at one-loop and beyond, 
and show how the LY theorem
(or its failure) can be understood in a very direct and simple way at the operator level.

\section{The Abelian case}
\label{sec:abelian}

It is instructive to first reconsider the original
Landau-Yang theorem, highlighting the difficulties in extending it 
to the case of a color-octet state. We denote  by
\be
M(1,2) \equiv\langle\gamma(k_1,\epsilon_1)\gamma(k_2,\epsilon_2)|V(P,\epsilon)\rangle\ ,
\ee
the amplitude for the  decay 
$V\to \gamma\gamma$, where  $V$ is a colorless massive spin-1 state and the 
$\gamma$'s are photons.  $k_1$ and $k_2$ are the  photons 4-momenta,
$P$ is the 4-momentum of $V$, and  momentum
conservation dictates $P=k_1+k_2 $. 
Relying only on Lorentz invariance and Bose symmetry ($M(1,2) = M(2,1)$), one can 
write
\ba
M &\sim& A\, (\epsilon_1 \cdot \epsilon_2) [\epsilon \cdot (k_1+k_2)]
\nonumber\\
&+& B\,[(\epsilon_1 \cdot \epsilon)(\epsilon_2\cdot k_1)
+ (\epsilon_2 \cdot \epsilon)(\epsilon_1\cdot k_2)]
\nonumber\\
&+&  C\,[\epsilon \cdot (k_1 + k_2)] (\epsilon_1 \cdot k_2)(\epsilon_2\cdot k_1)\ ,
\label{LYsinglet}
\ea
where $\epsilon_1\equiv\epsilon_1^*(k_1)$, $\epsilon_2\equiv\epsilon_2^*(k_2)$ and 
$\epsilon\equiv \epsilon(P)$ are the polarization 4-vectors
of the two photons and of the massive vector state respectively. $A,B,C$ are 
coefficients that can be determined in perturbation theory:
they only depend on scalar products of momenta, and therefore are constants
in the decay of a massive vector particle.  
Terms that vanish because of transversality,
$\epsilon_1\cdot k_1 = \epsilon_2\cdot k_2 = 0$, have not been
included. Instead, for the sake of introducing the discussion of the
non-Abelian case that will follow, we have kept in
Eq.~(\ref{LYsinglet}) also the $A$ and $C$ terms, even if in 
 the Abelian case these terms are trivially
equal to zero as a consequence of the transversality
of the massive vector polarization, $\epsilon (P) \cdot P=0$.

It is convenient to consider the $B$ term in Eq.~(\ref{LYsinglet}) after a Lorentz
transformation to the rest frame of the
massive state $V$. The photon polarizations transform under the
Lorentz transformation, yielding in general non vanishing time components
for $\epsilon_{1,2}^\mu$. However, the gauge invariance of the amplitude $M$
in Eq.~(\ref{LYsinglet}) is equivalent to the invariance of $M$ under the
 transformations \be \epsilon^{\mu}_j(k_j) \to \tilde
\epsilon^{\mu}_j(k_j)= \epsilon^{\mu}_j(k_j) + \beta_j k_j^\mu\
,\quad\quad (j=1,2)\ ,
\label{eq:poltransf}
\ee
where $\beta_j$ are arbitrary constants. This means that we are allowed to use $\tilde\epsilon_{1,2}$ in place of $\epsilon_{1,2}$ in the expression of $M$. In particular, we choose $\beta_j$ in such a way that
\be
\tilde\epsilon^{\mu}_j(k_j,\pm) =  \frac{1}{\sqrt{2}}(0,\mp 1,-i,0) \ .
\ee
Therefore, only polarizations transverse to the $\hat{\bm z}$ direction will eventually appear in $M$. 
Taking  the 3-momenta $\bm{k}_1$ and $\bm{k}_2$ along $\hat{\bm z}$,
in the rest frame of $V$, we simply have $\bm{k}_1 = -\bm{k}_2$ so that 
\begin{equation}
\tilde \epsilon_{1(2)}\cdot k_{2(1)} = 0 \, .
\label{prodmisto}
\end{equation}
$M$ in Eq.~(\ref{LYsinglet}) is therefore zero and $V\to\gamma\gamma$ is forbidden, leading to a proof of the Landau-Yang theorem in the Abelian case.\footnote{If the initial particle were a $1^+$ axial-vector, the only non-trivial parity conserving term would be proportional to
$\epsilon_{\mu\nu\rho\sigma}\epsilon^\mu \epsilon_1^\nu \epsilon_2^\rho (k_1-k_2)^\sigma$. In the rest frame of the massive vector particle 
$k_1-k_2=(0,0,0,2{\mathcal E})$, ${\mathcal E}$ being the photon energy. Thus $\epsilon,\epsilon_1,\epsilon_2$ can only have $0,1,2$ indices and permutations. 
Since $\epsilon_{1,2}$ are transverse polarizations, and $\epsilon^0\sim |\bm P|/M$ (where $\bm P$ and $M$ are the massive vector momentum and mass), we get $\epsilon_{\mu\nu\rho\sigma}\epsilon^\mu \epsilon_1^\nu \epsilon_2^\rho (k_1-k_2)^\sigma=0$. This illustrates 
that the Landau-Yang theorem applies as well to positive parity vectors. 
}
 The treatment of gauge invariance in the non-Abelian case, discussed in the next section, will require instead more care.

Before considering the non-Abelian case, 
it is instructive to understand the Landau-Yang result at the
  level of operators in the Lagrangian (or effective action) that describes
  the interaction between a color-singlet massive vector $V^\mu$ and the electromagnetic
  tensor $F^{\mu\nu}$.
  This
  approach takes care of gauge invariance automatically and further
  cancellations can be shown by using the field equations of motion,
  as illustrated below. The only non-trivial
  operator\footnote{Operators like
    $(\partial^\mu V_\mu) F_{\nu\rho}F^{\rho\nu}$ are trivial in the
    sense that they can be removed by using $\partial^\mu V_\mu=0$. }
  leading to the relevant Lorentz structures for the
  $V\rightarrow \gamma\gamma$ decay written in Eq.~(\ref{LYsinglet})
  is: \be \Delta {\cal L} = a (\partial^\mu V_\nu)
  F_{\mu\rho}F^{\rho\nu}\ ,
\label{LYop}
\ee
where $a$ is some coefficient with dimension mass$^{-2}$. 
Other operators with more derivatives just modify the 
momentum-dependent form factors of the vertices without changing its Lorentz structure.

Integrating by parts judiciously,  the operator in Eq.~(\ref{LYop}) can be rewritten as
\be
\Delta {\cal L}  = - a V_\nu (\partial^\mu F_{\mu\rho})F^{\rho\nu}-\frac{a}{4}(\partial^\nu V_\nu) F_{\mu\rho}F^{\mu\rho}\ .
\label{LYopi}
\ee
The second term, which can be written in this form thanks to the symmetry of $F_{\mu\rho}F^{\mu\rho}$ under photon exchange, can be dropped due to $\partial^\nu V_\nu=0$.
In the first, we can use the equation of motion of the electromagnetic field
to replace $\partial^\mu F_{\mu\rho}$ by a sum over all electromagnetic currents to which the photon couples in the theory. This replacement, equivalent to a field redefinition, does not change the physics \cite{EoM}, but makes clear the fact that this Lagrangian does not contribute to the $V\rightarrow \gamma\gamma$ decay with on-shell photons. (The modified Lagrangian in terms of electromagnetic currents will give the right description for processes with virtual photons coupled to such currents in the final state.)

\section{Non-Abelian case: Tree-level cancellation}
\label{sec:tree-level}

As mentioned in the Introduction  the $V^a\to g^bg^c$ amplitude 
is known to vanish at tree level. 
The purpose of this section is to trace the origin of the tree-level cancellation,
showing it explicitly.

We choose to work with a colored vector field $V^a$ rather than in full QCD and 
projecting a quark-antiquark pair onto a spin-one color-octet state.
We therefore introduce a pure-glue SU($N$) Lagrangian, to which we add a massive
colored vector field $V^a_\mu$, in the adjoint representation of
SU($N$), interacting with the gluons in a gauge invariant way:
\be
{\cal L}_\mathrm{VF} =  -\frac{1}{4} V_{\mu\nu}^a  V^{a\mu\nu} 
- \frac{1}{2}M^2 V_\mu^a V^{a\mu}
-\frac{1}{4} F_{\mu\nu}^a  F^{a\mu\nu} + {\cal L}_\mathrm{I}\ .
\label{eq:lagrangian}
\ee
Considering operators of dimension $d=4$, there is only one
operator that can contribute to the interactions: 
\ba
{\cal L}_\mathrm{I} &=& \label{eq:gaugeint}
\frac{g'}{2}  V_{\mu\nu}^a  F^{a\mu\nu} \\
&=& g'\partial_\mu V_\nu^a F^{a\mu\nu} - 
g g^\prime f^{abc}A_\mu^cV_\nu^b(\partial^\mu A^{a\nu}-\partial^\nu
A^{a\mu})\nonumber \, .
\ea
In this expression
\be
V_{\mu\nu}^a \equiv D_\mu^{ab} V_\nu^b-D_\nu^{ab} V_\mu^b \, ,
\ee 
$D_\mu^{ab}$ denotes the covariant derivative in the adjoint representation
\be
D_\mu^{ab}=\partial_\mu \delta^{ab}-gf^{abc}A_\mu^c \, ,
\ee
and
\be
F_{\mu\nu}^a = \partial_\mu A_\nu^a - \partial_\nu A_\mu^a + g f^{abc} A_\mu^b
A_\nu^c \, .
\ee

\begin{figure}
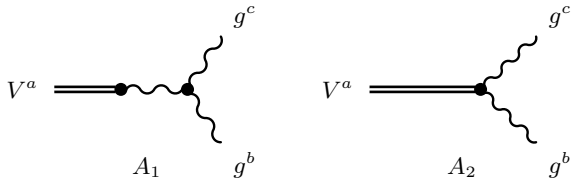

 \begin{center}
\begin{minipage}{3cm}
\include{Vgg3}
\end{minipage}
~~~~~~~~~
\begin{minipage}{3cm}
\include{Vgg2}
\end{minipage}
 \end{center}
\caption{\small Polar $(A_1)$ and direct $(A_2)$ contributions to the $V^a\to g^bg^c$ amplitude.\label{unic}}
\end{figure}

Using the interaction term between $V$ and the gluons introduced in  
Eq.~(\ref{eq:gaugeint}) we can extract the relevant contributions from the
 two vertices directly relevant to our calculation,
$V\to g$ and  $V \to gg$, and use them to evaluate the two amplitudes in
Fig.~\ref{unic}.

From the first term in the r.h.s. of  Eq.~(\ref{eq:gaugeint}) we get, after integration by 
parts~\footnote{We 
omitted null terms like $\partial^\nu V_\nu^a=0$ and terms proportional to 
$k_1\cdot \epsilon_2$ and $k_2\cdot \epsilon_1$ that do not contribute to the 
final amplitude because of the general transversality considerations made in Section~\ref{sec:abelian}.}
\be
V^a\to g^a :\quad - g^\prime\, M^2\epsilon_\nu \, ,
\label{eq:vertexVg}
\ee
where $M$ is the mass of $V$, whereas from the second term we get
\be
V^a\to g^bg^c :\quad i g g^\prime f^{abc}(\epsilon_2\cdot \epsilon_1) [ \epsilon\cdot (k_1-k_2)] \ .
\label{eq:vertexVgg}
\ee
In both cases, polarization vectors for legs that will eventually be external
ones have already been included.

In order to complete the amplitude $A_1$ in  Fig.~\ref{unic} 
we need to propagate the gluon with $1/(k_1 + k_2)^2 = 1/M^2$  from the $V$ state 
to the  triple-gluon vertex.
This leads to 
\be
A_1 = - ig g^\prime M^2\epsilon_\nu\,\frac{1}{M^2}\,  f^{abc} (k_1-k_2)^\nu(\epsilon_1\cdot 
\epsilon_2)\ .
\label{eq:amp1}
\ee
The amplitude $A_2$ is instead given simply by the expression 
for the $V\to gg$ vertex in Eq.~(\ref{eq:vertexVgg}). The sum $A_1+A_2$ shows a
full cancellation.

In Ref.~\cite{Cho:1995vh} the same conclusion was reached by computing explicitly the
$Q\bar Q\to gg$ amplitude with off-shell gluons after projecting the $Q\bar Q$ pair onto an $\ell=0$ (i.e. $S$-wave), 
spin 1, color-octet state, and later setting $k_1^2 = k_2^2 = 0$.
The result found in Ref.~\cite{Cho:1995vh} with off-shell gluons has the form 
\ba
M^{bca} &\sim&
f^{bca}\left\{
D\, (\epsilon_1 \cdot \epsilon_2) [\epsilon \cdot (k_1 - k_2)]\right.
\nonumber\\
&+& \left. E\,[(\epsilon_1 \cdot \epsilon)(\epsilon_2\cdot k_1)
- (\epsilon_2 \cdot \epsilon)(\epsilon_1\cdot k_2)]\right\}
\label{ChoLeibovich}
\ea
with 
\begin{equation}
D=E/2= 1+\frac{M^2}{k_1^2+k_2^2-M^2} \, .
\end{equation}
In this expression the second term comes from the sum of the two amplitudes with a propagating 
quark, whereas the first one originates from the amplitude with the tri-linear gluon vertex.  
In the on-shell limit $k_1^2=k_2^2=0$ and one finds $D=E=0$.

We have therefore shown that the vanishing of the amplitude for the decay of a massive
color-octet vector bosons into two massless gluon is due to a cancellation between 
the diagram with the triple-gluon vertex and the rest of the amplitude, whose forms are 
themselves dictated by gauge invariance.
When working in
full QCD and with a projected $Q\bar Q$ pair instead of the vector boson $V$, the cancellation
takes place between the antisymmetric combination of the Abelian diagrams and the
triple-gluon vertex diagram.

As we will show below, this delicate cancellation does not survive quantum corrections.

\section{Non-Abelian case: full analysis}
\label{sec:non-abelian}

We consider now the process where a massive color-octet vector state decays
into two massless colored gluons, $V^a\to g^b g^c$, in full generality. 
Lorentz invariance and Bose symmetry lead to the matrix element
\ba
M^{bca} &\sim& \epsilon^{\rho} \epsilon_1^{\mu}  \epsilon_2^{\nu} A_{\mu\nu\rho}^{bca} \nonumber\\
&=&
f^{bca}\left\{\frac{}{}
D\, (\epsilon_1 \cdot \epsilon_2) [\epsilon \cdot (k_1 - k_2)]\right.
\nonumber\\
&+& \left. E\,[(\epsilon_1 \cdot \epsilon)(\epsilon_2\cdot k_1)
- (\epsilon_2 \cdot \epsilon)(\epsilon_1\cdot k_2)]\right.
\nonumber\\
&+& \left. \frac{F}{k_1\cdot k_2}\,[\epsilon \cdot (k_1 - k_2)] (\epsilon_1 \cdot k_2)(\epsilon_2\cdot k_1)
\right\}\ ,
\label{LYoctet}
\ea
where $f^{abc}$ are the color-SU($N$) structure constants.
We have now denoted by $\epsilon_{1} \equiv \epsilon_1^{b*}(k_1)$, $\epsilon_2 \equiv
\epsilon_2^{c*}(k_2)$ and $\epsilon \equiv
\epsilon^{a*}(P)$ the polarizations of 
the massless colored gluons.  Note the minus sign between $k_1$ and $k_2$ momenta in the $D$ and $F$ terms, as well as the minus sign in the $E$ 
term,  compared to Eq.~(\ref{LYsinglet}).
Since $\epsilon\cdot (k_1-k_2)\neq 0$, the $D$ and $F$ terms cannot be dropped as was done in the Abelian case. 
An additional term proportional to the symmetric tensor $d^{bca}$, rather than to the
antisymmetric one $f^{bca}$, is identical to the two-photons case
considered previously, and will therefore not be discussed further.

In the following we exploit the BRST symmetry~\cite{BRST} of the gauge-fixed action  to constrain the form of the amplitude in Eq.~(\ref{LYoctet}). We will show that, differently from the Abelian case, we cannot conclude that all its terms are simultaneously zero.

The tensor
$A^{bca}_{\mu\nu\rho}(k_1,k_2)$ can be readily extracted from
Eq.~(\ref{LYoctet}):
\ba A^{bca}_{\mu\nu\rho} (k_1,k_2) &=& f^{bca} \left[\frac{}{}
  D g_{\mu\nu} (k_1 - k_2)_\rho \right. \nonumber\\
&+& \left.
  E \,(g_{\mu\rho} k_{1\nu} - g_{\nu\rho} k_{2\mu})\right. \nonumber\\
&+& \left. \frac{F}{k_1\cdot k_2}\, (k_1 - k_2)_\rho \, k_{2\mu} k_{1\nu}\right] \ .
\label{eq:lorentz}
\ea
As shown in the Appendix, BRST invariance implies that
this tensor satisfies
\be
\label{eq:transv}
k_1^\mu \epsilon_2^{\nu} A^{bca}_{\mu\nu\rho}(k_1,k_2) = 
\epsilon_1^{\mu} k_2^\nu A^{bca}_{\mu\nu\rho}(k_1,k_2) = 0
\, ,
\ee
which, together with Eq.~(\ref{eq:lorentz}), gives
\ba
0 &=& k_1^\mu \epsilon^{\nu}_2 A^{bca}_{\mu\nu\rho} \nonumber \\
&=& f^{bca} \left[ (D+E+ F) (k_1\cdot\epsilon_2)  k_{1\rho}
\right. \nonumber \\ 
&-&\left.  (D+F) (k_{1}\cdot\epsilon_2) k_{2\rho} 
-  E  (k_1\cdot k_2) \epsilon_{2\rho}  \right] \, ,
\label{eq:transv2}
\ea
where we set $k_1^2 = 0$, i.e. on-shell gluon, and used the transversality 
condition $k_2\cdot \epsilon_2 = 0$.
From Eq.~\ref{eq:transv2} we conclude that BRST invariance requires
$E=0$ and $D+F=0$. However the latter relation can be
satisfied even if $D\neq 0$ and $F\neq 0$.  

As discussed in the Abelian case, once BRST invariance is enforced, we
can work with transverse polarizations and hence we know that the $E$ and
$F$ terms will not contribute to the amplitude because terms of the form $\epsilon_1\cdot k_2$ and $\epsilon_2\cdot k_1$ vanish in an appropriate frame.
However, we  also see that the $D$ term can survive in the non-Abelian case,
consistently with the non-vanishing results at one loop obtained in Refs.~\cite{Chivukula:2013xla,Ma:2014oha}

The explicit tree-level calculation performed in Section~\ref{sec:tree-level} shows the cancellation of the $V^a$ gluonic decay amplitude expected from (\ref{eq:gaugeint}). 
As done above for the Abelian case, it is interesting to understand this
  result directly in terms of the operators that can appear in the Lagrangian. 
The $d=4$ operator in (\ref{eq:gaugeint}) generates the Lorentz structures  of the terms $E$ and $D$ in Eq.~(\ref{LYoctet}). By themselves, these terms do not contribute to the $V^a\to g^bg^c$ decay, as we have seen. At the operator level, that result is immediate to see in analogy to what we did for the Abelian case.  Simple integration by parts yields
\be
\frac{g'}{2}V^a_{\mu\nu}F^{a\, \mu\nu}=-g'V_\nu^b(D_\mu^{ba}
F^{a\,\mu\nu})\ ,
\ee 
and then, use of the equation of motion for the gluons to replace
$D_\mu^{ba} F^{a\mu\nu}$ by colored quark currents shows that the decay amplitude into on-shell gluons is zero.

The operator that can generate both $D$ and $F$ terms (in the combination
$D+F$), and thus contribute to the gluonic decay of $V^a$, is
\be
\Delta{\cal L}_6= 
f^{bca}V_{\mu\nu}^aF^{b\,\nu\rho}F^{c\,\mu}_\rho\ ,
\ee
a $d=6$ operator that can be radiatively generated as a finite correction to the effective action already at 1-loop order. Consistently with the results of our previous discussion, attempts to reduce this operator to equations of motion or trivially vanishing terms, as we did for the Abelian operator (\ref{LYop}), fail
in this case due to the non-Abelian nature of the gauge symmetry: no expression like (\ref{LYopi}) exists in this case. 
Note that the additional $d=6$ operator, $D^{ab\, \mu} V^b_{\mu\nu}D^{ac}_\rho F^{c\, \rho\nu}$ will not contribute
to the two-gluon decay of $V^a$, as can be shown by direct use of the gluonic equation of motion. The relevance of these $d=6$ operators for technicolor phenomenology and this particular gluonic decay amplitude has been studied in \cite{Chivukula,Bai:2014fkl}.

\section{Landau-Yang in $\ell=1$ scattering states}
Up to this point we have considered the decay of a colored massive vector particle.  
Now we  wish to analyze the Landau-Yang selection rule for the $Q(p_1)\bar Q(p_2)\to g(k_1)g(k_2)$ annihilation in $P$-wave, 
with quarks treated as spinless for simplicity. This configuration constitutes a stand-in for that of spin-1/2 quarks whose projection
onto a given spin and angular momentum state leads to a $J=1$ vector state for the $Q\bar Q$ pair.

We start by considering the amplitude
\begin{eqnarray}
M = \epsilon_{1\mu}\epsilon_{2\nu} A^{\mu \nu} (k_1,k_2,p_1)= \langle k_1,\epsilon_1; k_2,\epsilon_2 | p_1, p_2\rangle, \quad
\end{eqnarray}
with on shell particles. 
The $A^{\mu \nu}$ tensor can be decomposed in terms of form factors as
\begin{eqnarray}
A^{\mu \nu}(k_1,k_2,p_1) &=& A_1^{(\mp)}\,  (p^{\mu}_{1} k_2^{\nu} \pm p^{\nu}_{1} k_1^{\mu})+ \notag\\
&+& A_2^{(\mp)}\, (p^{\mu}_{1} k_1^{\nu} \pm p^{\nu}_{1} k_2^{\mu})+ \notag\\
&+& A_3^{(-)}\,(k_1\cdot k_2)\, g^{\mu\nu}+A_4^{(-)}\,  k_1^{\nu}
    k_2^{\mu}+ \notag\\
&+& A_5^{(-)}\,  k_1^{\mu}k_2^{\nu}\ ,
\label{ff}
\end{eqnarray}
where $A_i$ are functions of the Mandelstam variables $s=(p_1+p_2)^2$ and $t=(p_1-k_1)^2$.
The form factors $A_i^{(\pm)}$ are symmetric/antisymmetric with respect to $k_1\leftrightarrow k_2$ exchange. In Eq.~(\ref{ff}) we assume that the gluon configuration is color-odd so that  the final state is Bose-symmetric. 

BRST identities imply that if we saturate Eq.~(\ref{ff}) with $k_{1\mu} {\epsilon}_{2\nu}$ 
we get 
\begin{eqnarray}
 0&=&k_{1\mu} {\epsilon}_{2\nu}A^{\mu \nu}=\notag\\
&=& [A_2^{(-)} + A_2^{(+)}]\,  (k_1\cdot p_{1}) \,(k_1\cdot \epsilon_2 ) +\notag\\
&+& [A_2^{(-)} - A_2^{(+)}]\,  (k_1\cdot k_{2}) \,(p_1\cdot \epsilon_2 ) +\notag\\
&+& [A_3^{(-)} +A_4^{(-)} ](k_1\cdot k_2) (k_1\cdot \epsilon_2)\ .
\label{brsff}
\end{eqnarray}
From this equation we deduce that $A_2^{(-)} = A_2^{(+)} = 0$, while
due to the transversality of the gluon polarizations,
the $A_1^{(\mp)}$ and the $A_5^{(-)}$ contributions to the physical amplitude vanish.

One can further see that, in the threshold limit, $\bm p_1=\bm p_2=0$, we have $A_i^{(-)}=0$
because of antisymmetry, and therefore $A_{3,4}^{(-)}=0$ at threshold.
However this is not sufficient to imply the Landau-Yang selection rule. In fact 
the initial state $| p_1, p_2\rangle $ is to be projected onto the $P$-wave in the center of mass according to 
\begin{eqnarray}
| p_1, p_2; \ell=1,m \rangle = \int d \Omega_{\hat{\bm p}_1} Y^{\ell=1}_m (\hat{\bm p}_1) | p_1,p_2\rangle \ , 
\label{projection}
\end{eqnarray}
where $\bm p_1+\bm p_2=0$ and $\hat{\bm p}_1=\bm p_1/||\bm p_1||$ is the unit vector along $\bm p_1$, and the transition amplitude is given by
\be
M \propto \epsilon_{1\mu}\epsilon_{2\nu}\int d \Omega_{\hat{\bm p}_1} Y^{\ell=1}_m (\hat{\bm p}_1)\, A^{\mu \nu} \ .
\label{tram}
\ee
The $P$-wave condition selects  contributions to $A^{\mu \nu}$ that overall contain  one power of $p_{1}$ which originates from the $t$ dependence of  $A_i$'s which, close to threshold,  we parameterize as
\be
A_i^{(-)} \simeq  (\bm p_1\cdot \bm k_1) \, B_i\ ,
\ee
with $B_i$ constrained by~(\ref{brsff}). 
Replacing in Eq.~(\ref{tram}) we find
\begin{eqnarray}
M &\propto& B_3 \, \epsilon_{1\mu}\epsilon_{2\nu}\, g^{\mu\nu}\int d \Omega_{\hat{\bm p}_1} Y^{\ell=1}_m (\hat{\bm p}_1)\,  (\bm p_1\cdot \bm k_1) \propto\notag \\
&\propto &  B_3\, (\epsilon_{1}\cdot \epsilon_{2})\; k_1^{m}\ ,
\label{tramb}
\end{eqnarray}
where $\sqrt{2}\,k^{\pm 1}_1=k_{1x}\mp i k_{1y}$,  $k^{0}_1=k_{1z}$ and we used Eq.~(\ref{prodmisto}).

The result in Eq.~(\ref{tramb}) shows that the form of the amplitude for the process is very constrained. However, one cannot say anything about the value of $B_3$, because the BRST identity in Eq.~(\ref{brsff}) only allows one to conclude that the combination $B_3 + B_4$ must vanish. This is the same kind of roadblock that was met in Section~\ref{sec:non-abelian}, and it shows that the Landau-Yang selection rule cannot be extended to non-Abelian gluons also in the case  in which the initial state is not a single particle  but a scattering continuum.

\section{Conclusions}

In this note we have considered the process where a massive, color-octet vector state decays into two on-shell massless gluons. The well-known result that this amplitude vanishes at tree level in QCD has been shown in general terms to be due to a cancellation between the color-antisymmetric, Abelian part of the amplitude and its non-Abelian part.

Using considerations of Lorentz invariance and Bose symmetry we have also written down the most general expression for this amplitude, and employed BRST invariance to constrain its form.
We have shown that, at variance with the Abelian case, one cannot conclude that the amplitude vanishes to all orders. This is consistent with recent evidence that the amplitude does indeed not vanish at one-loop. We have explained this result in terms of
 the emergence of higher-dimension operators, radiatively generated in the effective action already at one-loop order.
In addition, we offer a novel way of understanding the Landau-Yang theorem (or its failure) directly in terms of the relevant operators, manipulated by integration by parts or using field equations of motion.
We have also considered in detail a situation where the massive vector state 
is given by the projection of a spinless quark-antiquark pair onto an $\ell = 1$ angular momentum state, reaching the same conclusions.
\newline

\noindent
\textbf{Acknowledgments.} 
A.D.P. thanks the CERN Theory group for hospitality while this work was performed.
M.C. is supported in part by the ILP LABEX (ANR-10-LABX-63) financed by French state funds managed by the ANR within the Investissements d'Avenir programme under reference ANR- 11-IDEX-0004-02. 
L.D.D. is supported by STFC, grant ST/L000458/1, and the Royal Society, Wolfson Research Merit Award, grant WM140078. 
The work of J.R.E. is supported by the grants FPA2013-44773-P and FPA2014-55613-P
from the Spanish Ministry for Economy and Competitivity (MINECO), by the grant 2014SGR1450 from the
Generalitat de Catalunya and by the Severo Ochoa excellence program of MINECO 
under the grant SO-2012-0234.
\newline

\noindent
\textbf{Note added.}
While our note was being finalized Ref.~\cite{Beenakker:2015mra} appeared, 
also discussing the failure of the Landau-Yang theorem in a non-Abelian gauge theory and reporting that an explicit calculation shows that the $(Q\bar Q)_{\bm 8}\to gg$ amplitude at one loop is non-vanishing. This paper also prompted a revised version of Ref.~\cite{Pleitez:2015cpa} that now features a discussion of the Landau-Yang theorem in QCD.

\section*{Appendix}
We now detail the derivation of the identities in Eq.~(\ref{eq:transv}).
The BRST transformations are defined by
\begin{eqnarray}
\delta_\mathrm{BRST} A^a_\mu &\equiv&  \theta\, \delta A^a_\mu = \theta\, ({D_\mu})^{ab} c^b\ , \\
\delta_\mathrm{BRST} B^a &\equiv& \theta\, \delta B^a = 0\ , \\
\delta_\mathrm{BRST} {\bar c}^a &\equiv& \theta\, \delta {\bar c}^a = \theta\, B^a \ ,\\
\delta_\mathrm{BRST} c^a &\equiv& \theta\,\delta c^a = - \frac {1}{2} \, g \, \theta\, f^{abc} c^b c^c \, ,
\end{eqnarray}
where $\theta$ is an anticommuting parameter and $B^a$ are Lagrange multipliers enforcing the gauge fixing conditions.
The theory that we are considering also contains a colored massive vector field $V^a_\mu (x)$, whose BRST variation is given by
\begin{eqnarray}
\delta V_\mu^a (x) = -\frac{g}{2} f^{abc} \, c^b \, V_\mu^c \, .
\end{eqnarray}
The Lagrangian density for the gauge-fixed theory, 
\begin{eqnarray}
{\cal L} = {\cal L}_\mathrm{VF} - \delta \, ({\bar c}^a \, \partial_\mu A^{a\mu} ) - \frac {1} {2 \alpha} B^a B^a \, ,
\end{eqnarray}
satisfies $\delta \, {\cal L} = 0$ and therefore
 the expectation value of any BRST variation vanishes, 
 \be
\langle\delta O \rangle = 0 \, ,
\ee
where $\langle...\rangle$ denotes the vacuum $T$-product. 

In particular, from
\begin{eqnarray}
&&\delta \, ({\bar c}^a (x) A^b_\nu (y) V^c_\rho (z)) = \nonumber \\
&&= B^a (x) A^b_\nu (y) V^c_\rho (z) + {\bar c}^a (x) (D_\nu)^{bd} c^d (y) V^c_\rho (z) + \nonumber \\
&&- \frac {g}{2} f^{cde} {\bar c}^a (x) A^b_\nu (y) c^d (z) V^e_\rho (z)\ ,
\end{eqnarray}
we have
\begin{eqnarray}
&& 0 = \langle  T (B^a (x) A^b_\nu (y) V^c_\rho (z))  \rangle + \nonumber \\
&&+ \langle  T ({\bar c}^a (x) (D_\nu)^{bd} c^d (y) V^c_\rho (z))  \rangle + \nonumber \\
&&- \frac {g}{2} f^{cde} \langle  T ({\bar c}^a (x) A^b_\nu (y) c^d (z) V^e_\rho (z)) \rangle  \, .
\end{eqnarray}
Integrating over the Nakanishi-Lautrup field $B^a$ yields:
\begin{eqnarray}
&&\langle  T (B^a (x) A^b_\nu (y) V^c_\rho (z))  \rangle = \nonumber \\
&&=\alpha \, \partial^\mu_x \langle T (A^a_\mu (x) A^b_\nu (y) V^c_\rho (z)) \rangle \, ,
\end{eqnarray}
hence
\begin{eqnarray}
&&\alpha \, \partial^\mu_x \langle  T (A^a_\mu (x) A^b_\nu (y) V^c_\rho (z))  \rangle = \nonumber \\
&&= - \langle  T ({\bar c}^a (x) (D_\nu)^{bd} c^d (y) V^c_\rho (z))  \rangle + \nonumber \\
&&+ \frac {g}{2} f^{cde} \langle  T ({\bar c}^a (x) A^b_\nu (y) c^d (z) V^e_\rho (z))  \rangle  \, .
\label{identity}
\end{eqnarray}

The tensor $A^{bca}_{\mu\nu\rho}$ in Eq.~(\ref{LYoctet}) is given by the LSZ reduction formalism in the form
\begin{eqnarray}
&&A^{bca}_{\mu\nu\rho} (k_1,k_2) \propto  \nonumber \\
&&\propto \lim_{(k_1+k_2)^2 \rightarrow M^2} \lim_{k_1^2 \rightarrow 0} \lim_{k_2^2 \rightarrow 0} k_1^2 k_2^2 [(k_1 + k_2)^2 - M^2] \times \nonumber \\
&&\int dx dy \, \langle T (A^a_\mu (x) A^b_\nu (y) V^c_\rho (0)) \rangle e^{ik_1 x} e^{ik_2 y}  \, .
\label{MDEF}
\end{eqnarray}
Eqs.~(\ref{identity}) and (\ref{MDEF}) imply
\begin{eqnarray}
&&{k_1}^\mu A_{\mu \nu \rho}^{abc}  (k_1,k_2)  \propto  \nonumber \\
&&\propto \lim_{(k_1+k_2)^2 \rightarrow M^2} \lim_{k_1^2 \rightarrow 0} \lim_{k_2^2 \rightarrow 0} k_1^2 k_2^2 [(k_1 + k_2)^2 - M^2] \times \nonumber \\
&&\int dx dy\, \langle  T ({\bar c}^a (x) (D_\nu)^{bd} c^d (y) V^c_\rho (0)) \rangle e^{ik_1 x} e^{ik_2 y}  \, ,
\label{identity12}
\end{eqnarray}
where the second term in the r.h.s. of Eq.~(\ref{identity}) does not contribute because there are no single particle poles  in the channel of the composite operator $f^{cde} c^d (z) V^e_\rho (z)$.  As for the first term, the massless ghost can contribute a term proportional to the momentum ${k_2}_\nu$ in the channel of the operator $(D_\nu)^{bd} c^d (y)$. Eq.~(\ref{identity12}) can therefore be written as 
\begin{eqnarray}
{k_1}^\mu A_{\mu \nu \rho}^{abc} (k_1,k_2) \propto {k_2}_\nu {\cal A}_{\rho}^{abc} \, ,
\end{eqnarray}
so that
\begin{eqnarray}
{k_1}^\mu {\epsilon_2}^\nu \epsilon^\rho A_{\mu \nu \rho}^{abc}  (k_1,k_2) = ({k_2} \cdot {\epsilon_2}) \epsilon^\rho {\cal A}_{\rho}^{abc}  = 0 \, ,
\end{eqnarray}
because $k_2\cdot\epsilon_2=0$.
This proves Eq.~(\ref{eq:transv}).


\end{document}